\documentclass[a4paper,fleqn,usenatbib,letters]{mnras}

\usepackage{newtxtext,newtxmath}

\usepackage[T1]{fontenc}
\usepackage{ae,aecompl}

\usepackage{graphicx}	
\usepackage{amsmath}	
\usepackage{amssymb}	

\def\kpc{{\rm\,kpc}}
\def\kms{{\rm\,km\,s^{-1}}}
\def\msun{{\rm\,M_\odot}}

\def\etal{{\it et al.\ }}

\def\ie{{ i.e.,\ }}
\def\lta{\mathrel{\spose{\lower 3pt\hbox{$\mathchar"218$}}
     \raise 2.0pt\hbox{$\mathchar"13C$}}}
\def\gta{\mathrel{\spose{\lower 3pt\hbox{$\mathchar"218$}}
     \raise 2.0pt\hbox{$\mathchar"13E$}}}
     
     \def\Gyr{{\rm\,Gyr}}

\def\FeH{{\rm[Fe/H]}}

\usepackage{tikz}

\title[The Pristine survey X]{The Pristine survey X:  a large population of low-metallicity stars permeates the Galactic disk.}

\author[Federico Sestito et al.]{
Federico Sestito,$^{1,2}$\thanks{E-mail: federico.sestito@astro.unistra.fr}
Nicolas F. Martin,$^{1,3}$ Else Starkenburg,$^{2}$ Anke Arentsen,$^{2}$  
\newauthor Rodrigo A. Ibata,$^{1}$ Nicolas Longeard,$^{4}$ Collin Kielty,$^{5}$ Kristopher Youakim,$^{2}$ Kim A. Venn,$^{5}$ 
\newauthor David S. Aguado,$^{6}$ Raymond G. Carlberg,$^{7}$ Jonay I. Gonz\'alez Hern\'andez,$^{8,9}$ Vanessa Hill,$^{10}$
\newauthor  Pascale Jablonka,$^{4,11}$ Georges Kordopatis,$^{10}$ Khyati Malhan,$^{12}$ Julio F. Navarro,$^{5}$ \newauthor Rub\'en S\'anchez-Janssen,$^{13}$ Guillame Thomas,$^{14}$ Eline Tolstoy,$^{15}$ Thomas G. Wilson,$^{16,17}$ 
\newauthor Pedro Alonso Palicio,$^{10,8,9}$ Spencer Bialek,$^{5}$ Rafael Garcia-Dias,$^{8}$ Romain Lucchesi,$^{4}$ 
\newauthor Pierre North,$^{4}$  Yeisson Osorio,$^{8}$  Lee R. Patrick$^{8}$ and Luis Peralta de Arriba$^{16,6}$
\\
$^{1}$ Universit\'e de Strasbourg, CNRS, Observatoire astronomique de Strasbourg, UMR 7550, F-67000, France\\
$^{2}$ Leibniz Institute for Astrophysics Potsdam (AIP), An der Sternwarte 16, D-14482 Potsdam, Germany\\
$^{3}$  Max-Planck-Institut f\"ur Astronomie, K\"onigstuhl 17, D-69117, Heidelberg, Germany\\
$^{4}$ Institute of Physics, Laboratoire d'astrophysique, \'Ecole Polytechnique F\'ed\'erale de Lausanne (EPFL), Observatoire, CH-1290 Versoix, Switzerland\\
$^{5}$ Department of Physics and Astronomy, University of Victoria, PO Box 3055, STN CSC, Victoria BC V8W 3P6, Canada\\
$^{6}$ Institute of Astronomy, University of Cambridge, Madingley Road, Cambridge CB3 0HA, UK\\
$^{7}$ Department of Astronomy and Astrophysics, University of Toronto, Toronto, ON M5S 3H4, Canada\\
$^{8}$ Instituto de Astrof\'isica de Canarias, V\'ia L\'actea, 38205 La Laguna, Tenerife, Spain\\
$^{9}$ Universidad de La Laguna, Departamento de Astrof\'isica, 38206 La Laguna, Tenerife, Spain\\
$^{10}$ Laboratoire Lagrange, Universit\'e de Nice Sophia-Antipolis, Observatoire de la C\^{o}te d'Azur, CNRS, Bd de l'Observatoire, CS 34229, 06304 Nice cedex 4, France\\
$^{11}$ GEPI, Observatoire de Paris, Universit\'e PSL, CNRS, Place Jules Janssen, F-92190 Meudon, France\\
$^{12}$ The Oskar Klein Centre for Cosmoparticle Physics, Department of Physics, Stockholm University, AlbaNova, Stockholm, Sweden\\
$^{13}$ UK Astronomy Technology Centre, Royal Observatory, Blackford Hill, Edinburgh, EH9 3HJ, UK\\
$^{14}$ NRC Herzberg Astronomy and Astrophysics, 5071 West Saanich Road, Victoria, BC V9E 2E7, Canada\\
$^{15}$ Kapteyn Astronomical Institute, University of Groningen, Landleven 12, NL-9747AD Groningen, the Netherlands\\
$^{16}$ Isaac Newton Group of Telescopes, E-38700 Santa Cruz de La Palma, Spain\\
$^{17}$ SUPA, School of Physics and Astronomy, University of St. Andrews, North Haugh, Fife KY16 9SS, UK
}

\date{Accepted XXX. Received YYY; in original form ZZZ}

\pubyear{2019}

\begin{document}
\label{firstpage}
\pagerange{\pageref{firstpage}--\pageref{lastpage}}
\maketitle

\begin{abstract}
The orbits of the least chemically enriched stars open a window on the formation of our Galaxy when it was still in its infancy. The common picture is that these low-metallicity stars are distributed as an isotropic, pressure-supported component since these stars were either accreted from the early building blocks of the assembling Milky Way, or were later brought by the accretion of faint dwarf galaxies. Combining the metallicities and radial velocities from the Pristine and LAMOST surveys and Gaia DR2 parallaxes and proper motions for an unprecedented large and unbiased sample of very metal-poor stars at $\FeH\leq-2.5$ we show that this picture is incomplete. This sample shows strong statistical evidence (at the $5.0\sigma$ level) of asymmetry in their kinematics, favouring prograde motion. Moreover, we find that $31\%$ of the stars that currently reside in the disk do not venture outside of
the disk plane throughout their orbit. The discovery of this population implies that a significant fraction of stars with iron abundances $\FeH\leq-2.5$ formed within or concurrently with the Milky Way disk and that the history of the disk was quiet enough to allow them to retain their disk-like orbital properties.
\end{abstract}





\section{Introduction}

As successive generations of stars are formed from the gaseous material that is chemically enriched by earlier generations of stars, the most chemically pristine stars provide a unique window into the oldest components of the Milky Way  \citep{Freeman02,Karlsson13}, dating back to times when our Galaxy was still assembling. It is expected that low-metallicity stars, whose iron abundance is lower than a few thousandths of the Sun's ($\FeH\leq-2.5$) were formed at most 2--3$\Gyr$ after the Big Bang  \citep{ElBadry18}. Since the then proto-Milky Way was still in the process of chaotically accreting, it is commonly expected that the most metal-poor stars mainly trace the spheroid of the Milky Way (hereafter MW). These stars should either be present in the deepest parts of the Galactic potential well if they were accreted at the formation of the MW, or further out in the stellar halo if they formed in dwarf galaxies that were accreted onto the MW at later times  \citep{White00, Brook07, Gao10, Salvadori10, Tumlinson10, Ishiyama16, Starkenburg17b, ElBadry18, Griffen18}. The inescapable conclusion of this scenario is that low-metallicity stars should follow pressure-supported orbits and that they should be most prominent in the central regions of the MW or in its diffuse stellar halo. Moreover, these stars should be absent from the MW disk because stars formed very early in the proto-disk were scattered into the halo during the dynamic assembly process. The disk's successive generation of stars are expected to have formed from already enriched gas.

Recent work by \citet{Sestito19} has shown the orbital properties of the 42 most pristine stars known  in the ultra metal-poor regime (UMP,  $\FeH <-4.0$) using the photometric and kinematic data of the Data Release 2 (DR2) of the Gaia satellite. Surprisingly, roughly a quarter of those stars orbit close or within the plane of the MW disk. Whilst tentative, the small size of the sample and inhomogeneous data collection methods in this literature sample prevent a firm conclusion on the orbital parameters of the most metal-poor stars\footnote{Very similar kinematical signatures are found by \citet{DiMatteo19} for a small sample of 54 stars peaked around [Fe/H] = $-3$.}. In this work, we revisit these interesting findings with our more unbiased and very large sample of stars, putting the work on the orbital properties of very metal-poor stars on a much firmer statistical footing.

In general, the rarity of low-metallicity stars among the bulk of the more metal-rich MW stars has long limited the mapping of their distribution. However, recent, systematic, and large spectroscopic surveys  \citep{Allende14, Li18} or  specific photometric surveys  \citep{Wolf18,Starkenburg17a} yield increasingly large spectroscopic samples of such stars. In this work, we use two well-known samples in order to study the orbital properties of the most metal-poor tail of disk stars.
The Large sky Area Multi-Object fiber Spectroscopic Telescope (LAMOST) observed a sample of $\sim$5,000 stars at very low-metallicity ($\FeH<-2.0$). Moreover, these LAMOST stars probe all Galactic latitudes and were not selected to specifically focus on the regions of the MW halo at high Galactic latitudes. We complement this sample with $\sim$600 stars in the very metal-poor regime --- 67 of which are extremely metal-poor stars (EMP; with $\FeH<-3.0$ ) --- from the spectroscopic follow-up campaign of the Pristine survey  \citep{Youakim17, PristineDavid}. This survey is looking at the northern hemisphere, avoiding regions towards the MW disk, but it has the advantage that the initial candidates were selected from narrow-band photometry and therefore that the selection of targets is more easily tractable. The resulting combined sample of 1,027 stars below $\FeH \leq-2.5$ from LAMOST and the Pristine survey provides a unique dataset to study the orbital properties of very metal-poor stars, as it is both large and selected purely on metallicity without any pre-selection on kinematics. 

We describe the data samples in Section~\ref{data}, before turning to our results in Section~\ref{results} and implications for our understanding of the formation and (early) evolution of the MW galaxy in Section~\ref{discussions}.

\section{Data}\label{data}

\subsection{The Pristine sample}
The Pristine survey is a photometric survey that aims at efficiently finding the most metal-poor stars \citep{Starkenburg17a}. It is based on narrow-band Ca H$\&$K photometry obtained with the MegaCam wide-field camera on the 3.6m Canada-France-Hawaii Telescope. In this work, we use the very metal-poor stars (VMP, $\FeH <-2.0$) photometrically selected from the narrow-band photometry and then spectroscopically followed-up with the IDS spectrograph at the 2.54m Isaac Newton Telescope (INT) at Observatorio del Roque de los Muchachos. This sample and its analysis are described in \citet{PristineDavid}. The sample is composed of 583 genuine VMP stars, of which 67 are EMP stars ($\FeH <-3.0$), none are UMP. We derive the radial velocities of these VMP stars using the \texttt{fxcor} task (a Fourier cross-correlation method) from IRAF \citep{Tody86,Tody93} with an appropriate synthetic template spectra for each star matching within $250$ K in temperature, $0.5$ dex in $\FeH$, and 1 dex in carbon abundance. A sub-sample of these stars ($\sim 20$) was subsequently followed-up with high-resolution at CFHT with ESPaDOnS \citep{Kim19}  and at Gemini with GRACES (Kielty \etal, in prep.). From this overlapping sub-sample we assessed the magnitude of any systematic errors on the radial velocities and found a systematic offset of $\mu_{off}= 4.9\pm3.4 \kms$ in the mean and a standard deviation between both sets of measurements of $\sigma_{sys} = 10.5\pm4.1 \kms$. Together with the individual measurement uncertainties on the radial velocity derivation, these uncertainties are propagated in the derivation of the orbital parameters and their uncertainties.

\subsection{The LAMOST sample}
\citet{Li18} presented new metallicities for a set of 10,000 VMP star candidates from LAMOST DR3 \citep{Lamost12,Cui12}. We note there is a spurious effect in this VMP sample, where a lot of stars accumulate at the lower effective temperature limit of the employed model grid. Therefore, we clean this sample accordingly, resulting a final selection of 4838 VMP stars, of which 41 are EMP and none are UMP. For a detailed description of the cleaning steps see Appendix~\ref{A} and Figures~\ref{surveys}-\ref{lamost} therein.

\subsection{Determination of distances and orbital properties}
We infer distances for stars from both surveys following the Bayesian method described in \citet{Sestito19}. In short, we derive a probability distribution function (PDF) of the heliocentric distance to a star by combining its photometric (G, BP, and RP magnitudes) and astrometric data (parallax $\varpi$) from Gaia DR2 with a sensible MW stellar density prior and MESA/MIST isochrone models \citep{Dotter16, Choi16} for stars of old age ($>11$ Gyr). This Bayesian method to infer distance does not require a reliable parallax measurement, but does take into account all parallax information available (even negative values).
As discussed in \citet{Sestito19}, the choice of the MW density prior affects the results only when the distance PDF has two solutions (\ie both a dwarf and a giant solution) changing the probabilities associated to the two solutions, but not the values of the distances. After finding that a significant fraction of UMP stars reside close to the MW plane \citep{Sestito19}, we therefore chose a MW density prior composed by the sum of a halo component described by a power law, and a disk component described by an exponential distribution law. Subsequently, we derive the orbit using the Galpy code \citep{Bovy15} providing it with the inferred distances, the radial velocities and the exquisite Gaia DR2 proper motion. For the gravitational potential, we use a more massive halo ($1.2 \cdot 10^{12} \msun$) compared to \textit{MWPotential14} from Galpy ($0.8 \cdot 10^{12} \msun$) in agreement with the value from \citet{BlandHawthorn16}, an exponentially cut-off bulge, a Miyamoto Nagai Potential disc, and a \citet{NavarroFrenkWhite97} dark matter halo. The Local Standard of Rest circular velocity, Sun peculiar motion, and distance from the Galactic centre are the same assumed by \citet{Sestito19} (see also references therein). The table with the inferred orbital parameters is provided as Online material.

\section{Results}\label{results}

\begin{figure*}
\includegraphics[width=\hsize]{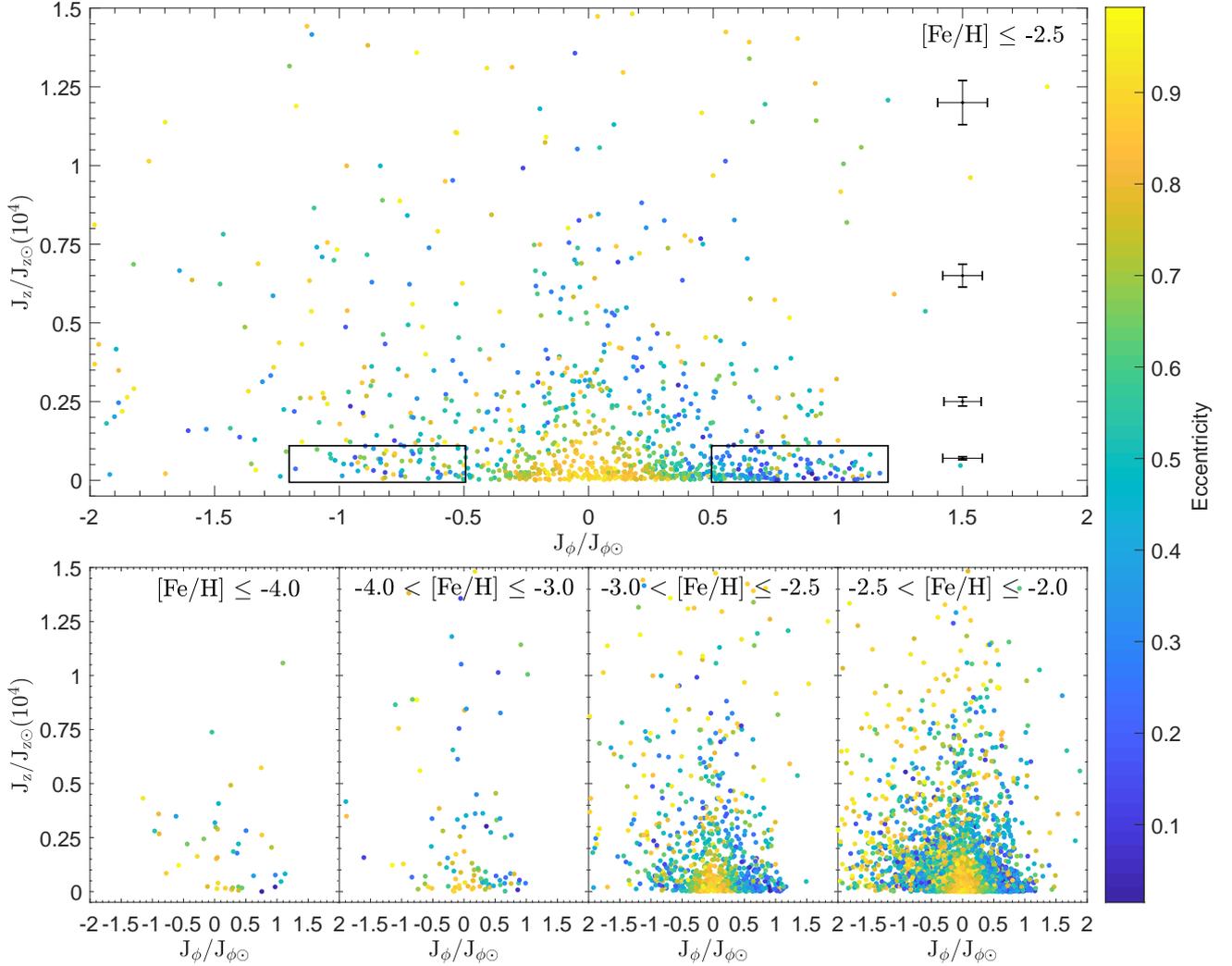}
\caption{Vertical action vs. azimuthal action component colour-coded by eccentricity. Top panel: our sample+\citet{Sestito19} stars with $\FeH \leq -2.5$ are shown. Typical uncertainties for four bins in $J_{z}/J_{z\odot}$ are shown on the right. Bottom left panel: UMP ($\FeH \leq -4.0$) stars from \citet{Sestito19}. Bottom centre-left panel: stars with $-4.0 < \FeH \leq -3.0$. Bottom centre-right panel:  stars with $-3.0 < \FeH \leq -2.5$. Bottom right panel: stars with $-2.5 < \FeH \leq -2.0$. The action quantities are scaled by the solar values (\ie $J_{\phi\odot}= 2009.92 \kms \kpc $, $J_{z\odot}= 0.35\kms \kpc$). We detect an asymmetry and the predominance for the prograde motion (right box in the top panel) vs. the retrograde planar stars (left box in the top panel) with $5.0\sigma$ level for stars with $\FeH\leq-2.5$ }\label{action}
\end{figure*}

We cross-identify our combined sample of low metallicity stars with the Gaia DR2 catalogue to infer the distances to the stars from the Gaia DR2 photometry and parallax measurements. We calculate the Galactic orbital properties from these distances, the radial velocities from the Pristine and LAMOST spectra, and the Gaia DR2 proper motions. In particular, we focus on three orbital properties: the azimuthal action $J_\phi$, which is equivalent to the z-axis component of a star's angular momentum; the vertical action, $J_z$, which conveys information about how far a star's orbit brings it away from the Galactic plane; and the eccentricity of the orbit, $\epsilon$. The top panel of Figure~\ref{action} shows the distribution of stars in the $J_z$--$J_\phi$ plane, colour-coded by the eccentricity of a given star's orbit, for our full sample with $\FeH\leq-2.5$, complemented by the 42 UMP stars ($\FeH \leq-4.0$)  from \citet{Sestito19}. We see a clear population of stars that remain close to the MW plane (very small $J_z$), although not all of these stars are on perfectly circular orbits. More importantly, the sample exhibits a strong asymmetry between prograde ($J_\phi>0$) and retrograde ($J_\phi<0$) stars, where prograde stars dominate with an angular momentum up to the Sun's value. 

The bottom panels of Figure~\ref{action} show the same action plot divided into 4 metallicity bins, respectively the UMP stars populated only by the 42 stars from \citet{Sestito19}, the $-4.0 <\FeH \leq-3.0$ bin, the $-3.0 <\FeH \leq-2.5$ regime, and, to be complete, the bin with $-2.5 <\FeH \leq-2.0$, where the signature of a disk population was already discovered \citep{Beers02,Reddy08,Ruchti11,Li17}. Separating the sample in these metallicity bins makes it evident that the prograde stars that remain close to the MW plane inhabit all $\FeH$ ranges. Focussing on the region of the diagram that is populated by disk-like stars, with $0.5<J_{\phi}/J_{\phi\odot}<1.2$ and $J_z/J_{z\odot}<0.125$, we assess the significance of the asymmetry through a direct comparison with the retrograde stars of similar properties ($-1.2<J_{\phi}/J_{\phi\odot}<-0.5$ and same $J_z/J_{z\odot}$ range). Assuming Poisson statistics, we find that the prograde region is $5.0\sigma$ overdense compared to its retrograde counterpart for the $\FeH\leq-2.5$ regime, or $1.9\sigma$ overdense for the $\FeH<-3.0$. For these two regimes, the overdensity of disk-like stars in the prograde box remains similar and the lower significance in the lower metallicity bin is driven by the smaller numbers. When adopting a two-dimensional Kolmogorov-Smirnov test  \citep{Peacock83,Fasano87}  we find that we can discard the hypothesis that these different samples, from the ultra metal-poor regime to the very metal-poor regime (VMP, $\FeH<-2.0$), are drawn from a different parent distribution. There is no bias we can think of in the two surveys at the base of our sample that would preferentially over-select prograde over retrograde stars. In particular, no selection on the motion of stars was apply to either of the two surveys that were designed before the Gaia DR2 data were available. We have tested that our results are similar whether we restrict ourselves only to these stars with reliable parallax information (see Appendix~\ref{B} and Figure~\ref{actionpar} therein).

In order to quantify the underlying fraction of disk-like stars in the low metallicity regime, we look at the population of low-metallicity stars located within $3 \kpc$ of the MW plane, from which we can identify two samples, (i) the disk-like stars with the maximum excursion from the MW plane $|Z_{max}|\le 3\kpc$ in a prograde motion ($J_{\phi}>0$) and (ii) the halo-like stars that are either passing through the disk or that are close to the plane in a retrograde motion. Of the population of stars with $\FeH\leq-2.5$ with $|Z|<3 \kpc$ $\sim31\%$ belongs to the disk-like sample (i). Although a disk-like component of the MW has been seen before down to $\FeH =-2.3$ \citep{Li17}, this is the first time we find strong evidence of such a population for the lowest metallicity stars ($\FeH\leq-2.5$). We conclude that an important fraction of the 1,069 low-metallicity stars from the sample in fact reside in the MW disk.

\section{Discussion and Conclusion}\label{discussions}
Understanding the origins of these stars has major implication for the assembly and evolution of the MW. Simulated disk galaxies for which maps are published of low-metallicity stars with $\FeH\leq-2.5$ in either density \citep{Tumlinson10,Starkenburg17b} or kinematical space \citep{ElBadry18} do not commonly bear this feature. This is either due to the MW having a unique formation path or to these simulations not including all the necessary physical ingredients to produce such a feature.
 We propose three different scenarios to explain this observational feature: minor mergers, the assembly of the proto-MW, and the in situ formation of this component of the disk at early times. We note that these scenarios are not mutually exclusive.

First, it is possible that the observed low metallicity stars were brought into the MW plane through the minor merging of small satellites that deposited their stars in the environment of the disk, that was already in place, after their orbit decayed via dynamical friction \citep{Scannapieco11} and the eccentricity enhanced by tidal interaction \citep{Abadi03,Penarrubia02}. Results from cosmological simulations have shown that the disrupted merged satellite can be aligned with the disk \citep{Gomez17}. Some simulations \citep{Scannapieco11,Karademir19} show that up to 5--20\% of the disk stars have not formed in situ but were brought in from now-merged satellites. 

Alternatively, or additionally, low-metallicity disk-like stars could have been born in and brought in from the building blocks that formed the disk of the proto-MW at early times. In such a scenario at high-redshift, we can expect that whatever gas-rich blocks formed the backbone of the MW disk also brought its own stars, including low-metallicity ones.

Cosmological simulations \citep{ElBadry18} show that, of all stars currently within $10 \kpc$ from the MW centre and formed before redshift $z = 5$, less than half were already in the main progenitor at $z = 5$. Over half of these extremely old stars would make their way into the main Galaxy in later merging events and find themselves at $z = 5$ inside different galaxies that are up to $250 \kpc$ away from the main progenitor centre. These two mergers scenarios can naturally funnel stars in the inner regions of the main galaxy, to be observed on orbits close to the disk plane today. 

For the third scenario, the in situ formation at early times, it is necessary to invoke the presence of pockets of pristine gas in the MW's gaseous disk during the first few Gyr of the Universe. This scenario implies that the MW plane was already defined within 2--3 Gyr and that this plane has not significantly changed over the last 10--11 Gyr. Consequently, the MW cannot have suffered dramatic merger and/or accretion events that would have likely tilted its disk and/or randomised the orbit of the EMP stars \citep{Scannapieco09}. Such a scenario would be in line with the commonly accepted idea that the MW has undergone a very quiet accretion history \citep{Wyse01,Stewart08}.  However, two main questions arise from this scenario. The first question is whether it is possible to form stars so completely devoid of metals in a relatively well-mixed interstellar medium disk in this stage of evolution of the MW.  The second question relates to the mechanisms that can push the stars from the small radius of their birth place to the solar neighbourhood and from the likely circular orbit of their birth to the range of observed eccentricities of the orbits we observe them on today. Radial migration is very efficient in pushing outwards the orbital radius conserving their circularity \citep{Sellwood02,Haywood08,Schonrich09}. For stars with higher orbital eccentricity at birth \citep{Brook04,Minchev13,Bird13}, radial migration is less efficient \citep{Martig14} but non-linear interactions between the MW bar and its spiral arms \citep{Minchev10} or perturbations from infalling minor mergers \citep{Quillen09} could redistribute their angular momentum.

One important implication of this work is that the disk region should not be avoided in the search and study for the most metal-poor stars, contrary to what has frequently been done in the past. Moreover, cosmological zoom-in simulations should be revisited to reproduce this population of low-metallicity stars with disk-like kinematics. Whatever the true origin of these prominent disk-like low metallicity stars, they undoubtedly open a window on the assembly of the oldest parts of the MW and pose a challenge to our understanding of very early Galaxy formation in general.

\section*{Acknowledgements}
We want to thank Ivan Minchev for the insightful comments and suggestions on radial migration. 
 This work has made use of data from the European Space Agency (ESA) mission {\it Gaia} (\url{https://www.cosmos.esa.int/gaia}), processed by the {\it Gaia} Data Processing and Analysis Consortium (DPAC, \url{https://www.cosmos.esa.int/web/gaia/dpac/consortium}). Funding for the DPAC has been provided by national institutions, in particular the institutions participating in the {\it Gaia} Multilateral Agreement. Guoshoujing Telescope (the Large sky Area Multi-Object fiber Spectroscopic Telescope, LAMOST) is a National Major Scientific Project built by the Chinese Academy of Sciences. LAMOST is operated and managed by the National Astronomical Observatories, Chinese Academy of Sciences. FS thanks the Initiative dExcellence IdEx from the University of Strasbourg and the Programme Doctoral International PDI for funding his Ph.D. This work has been published under the framework of the IdEx Unistra and benefits from a funding from the state managed by the French National Research Agency as part of the investments for the future program. FS, NFM, and RAI gratefully acknowledge support from the French National Research Agency (ANR) funded project ``Pristine" (ANR-18-CE31-0017) along with funding from CNRS/INSU through the Programme National Galaxies et Cosmologie and through the CNRS grant PICS07708. The authors acknowledge the support and funding of the International Space Science Institute (ISSI) for the international team ``Pristine". FS acknowledges the support and funding of the Erasmus+ programme of the European Union.




\bibliographystyle{mnras}
\bibliography{Refs} 




\appendix

\section{The cleaning of the LAMOST sample}\label{A}
\begin{figure*}
\includegraphics[width=\hsize]{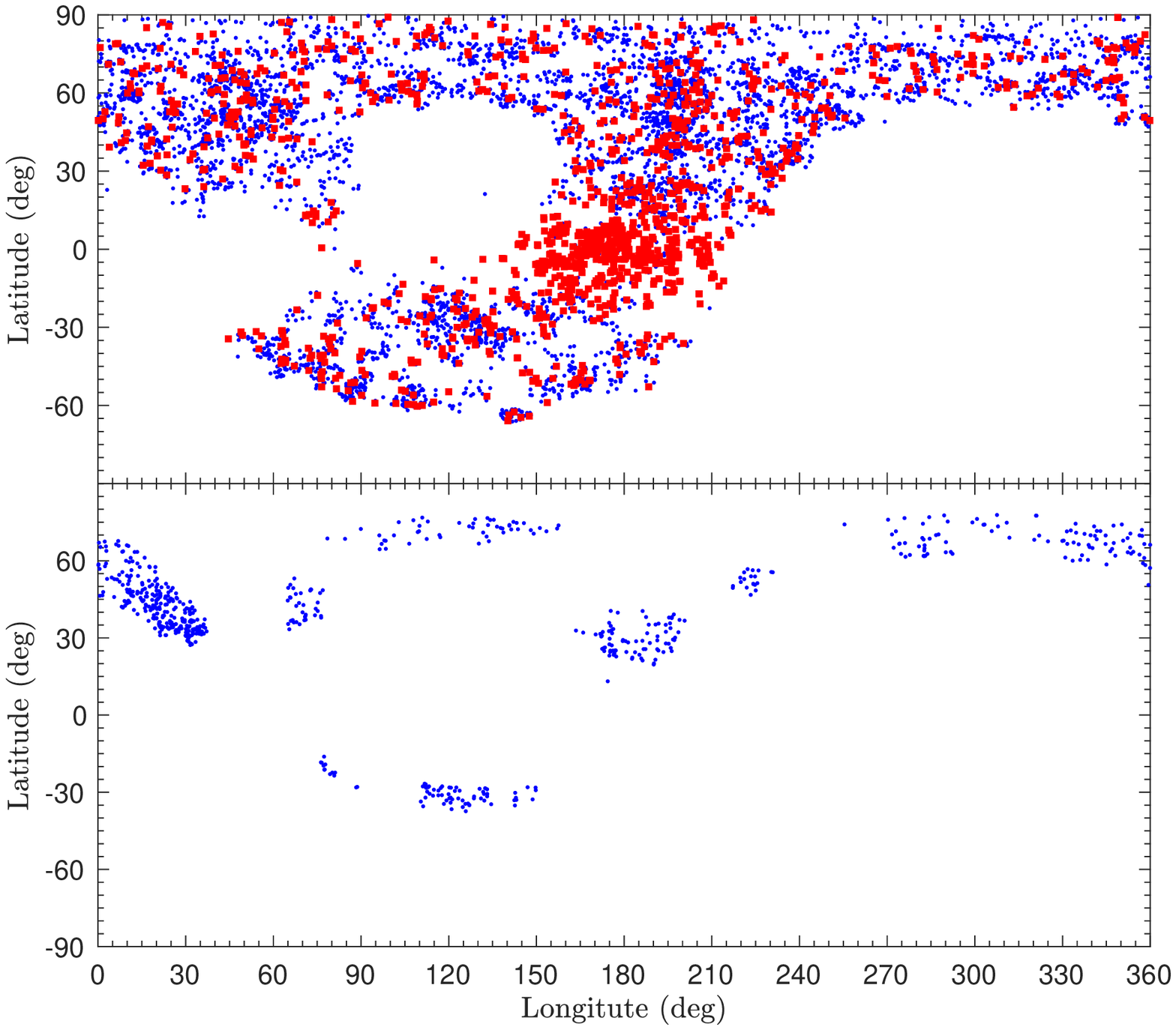}
\caption{Galactic longitude and latitude distribution of the Pristine and LAMOST \citep{Li18} sample. Top panel: the cleaned LAMOST (blue circles) and the removed stars from \citet{Li18} (red squares). The removed stars in this sample correspond to the stars with $T_{eff}<4500$ K, $\FeH_{Public}>-1.5$ and signal to noise ratio in g band $<20$ when $\FeH_{Li} <-3.0$. Bottom panel: the Pristine sample observed at INT. }\label{surveys}
\end{figure*}

\begin{figure*}
\includegraphics[width=\hsize]{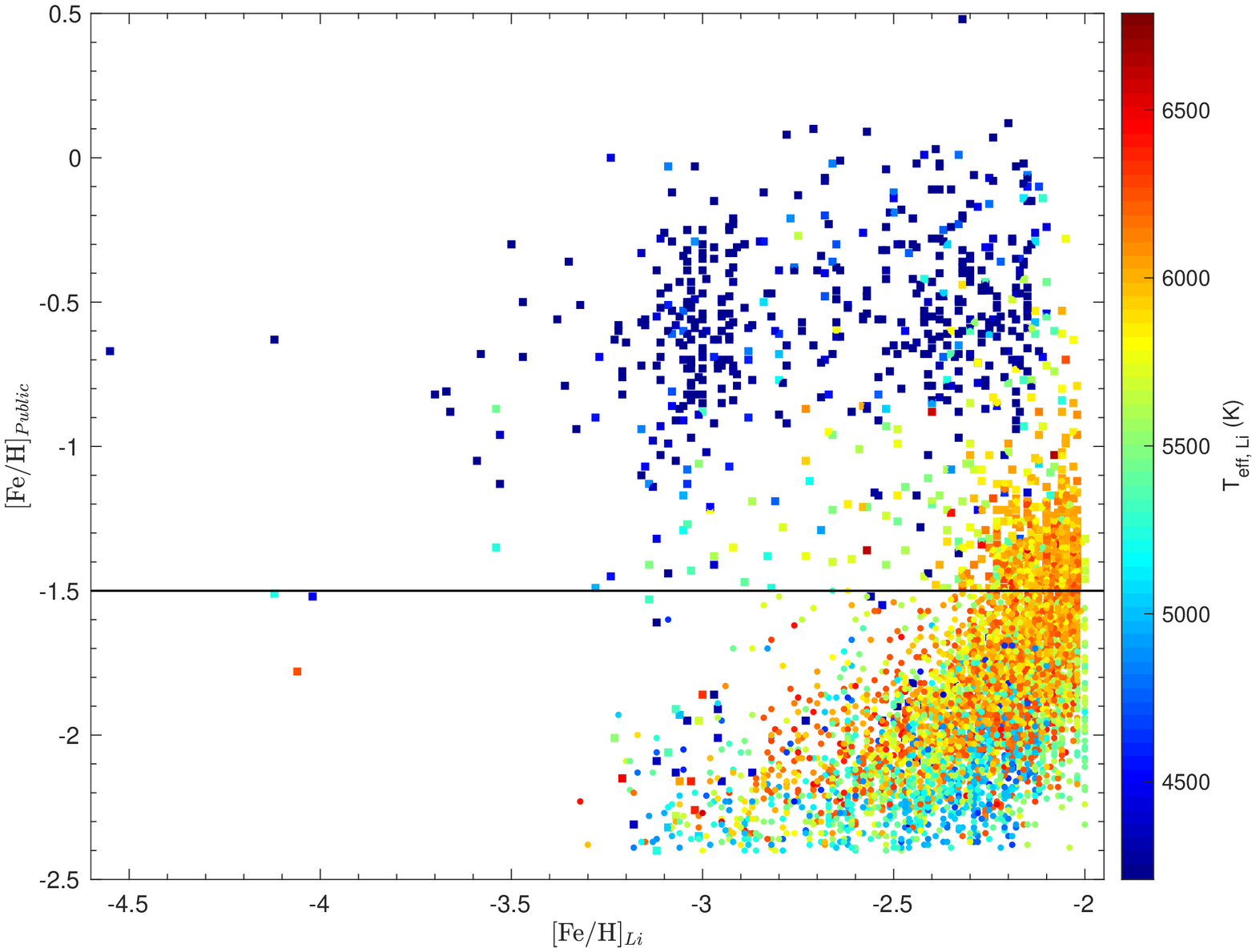}
\caption{Public LAMOST $\FeH$ vs. $\FeH$ from \citet{Li18} colour-coded  by effective temperature from \citet{Li18}. The cleaned sample (circle) is obtained by selecting stars with $T_{eff, Li} > 4500$ K, $\FeH_{Public} < -1.5$ and signal to noise ratio in the g band above 20 when $\FeH_{Li} <-3.0$. The removed stars are marked with a square. Note that the measurement of the $\FeH_{Public}$ saturates at the lower limit of $\FeH = -2.4$ (\ie a star with a true $\FeH = -3.0$ has a public LAMOST $\FeH=-2.4$).}\label{lamost}
\end{figure*}

There is a spurious effect in this VMP sample, where a lot of stars accumulate at the lower effective temperature limit of the employed model grid. Most of these stars are located in the disk plane, a region with high extinction (see Figure~\ref{surveys}). A cross-match with the public LAMOST DR3 \citep{Lamost12,Cui12} catalogue shows that many of these stars actually have high metallicities ($\FeH > -1$), see Figure~\ref{lamost}. An independent check of the metallicity of these stars with ULySS \citep{Koleva09,Arentsen19} confirms their metal-rich nature. We therefore clean the sample from contaminants ($\FeH > -2$) by selecting stars with $T_{eff, Li} > 4500$ K and removing stars with $\FeH_{Public} > -1.5$. We additionally require the signal-to-noise ratio in the blue part of the spectrum to be higher than 20 when $\FeH_{Li} <-3.0$ to ensure a robust determination of metallicity at these low metallicities. We note there is also an offset of $\sim0.5$ dex between the $\FeH$ from \citet{Li18} and the $\FeH$ from public LAMOST DR3 \citep{Lamost12,Cui12}, but since the $\FeH_{Li}$ values have been checked to be on the same scale as high-resolution observations \citep{Li18}, we have adopted these values. 
Our final selection results in a total of 4838 VMP stars, of which 41 are EMP and none are UMP.

\section{Results with good parallax data}\label{B}
\begin{figure*}
\includegraphics[width=\hsize]{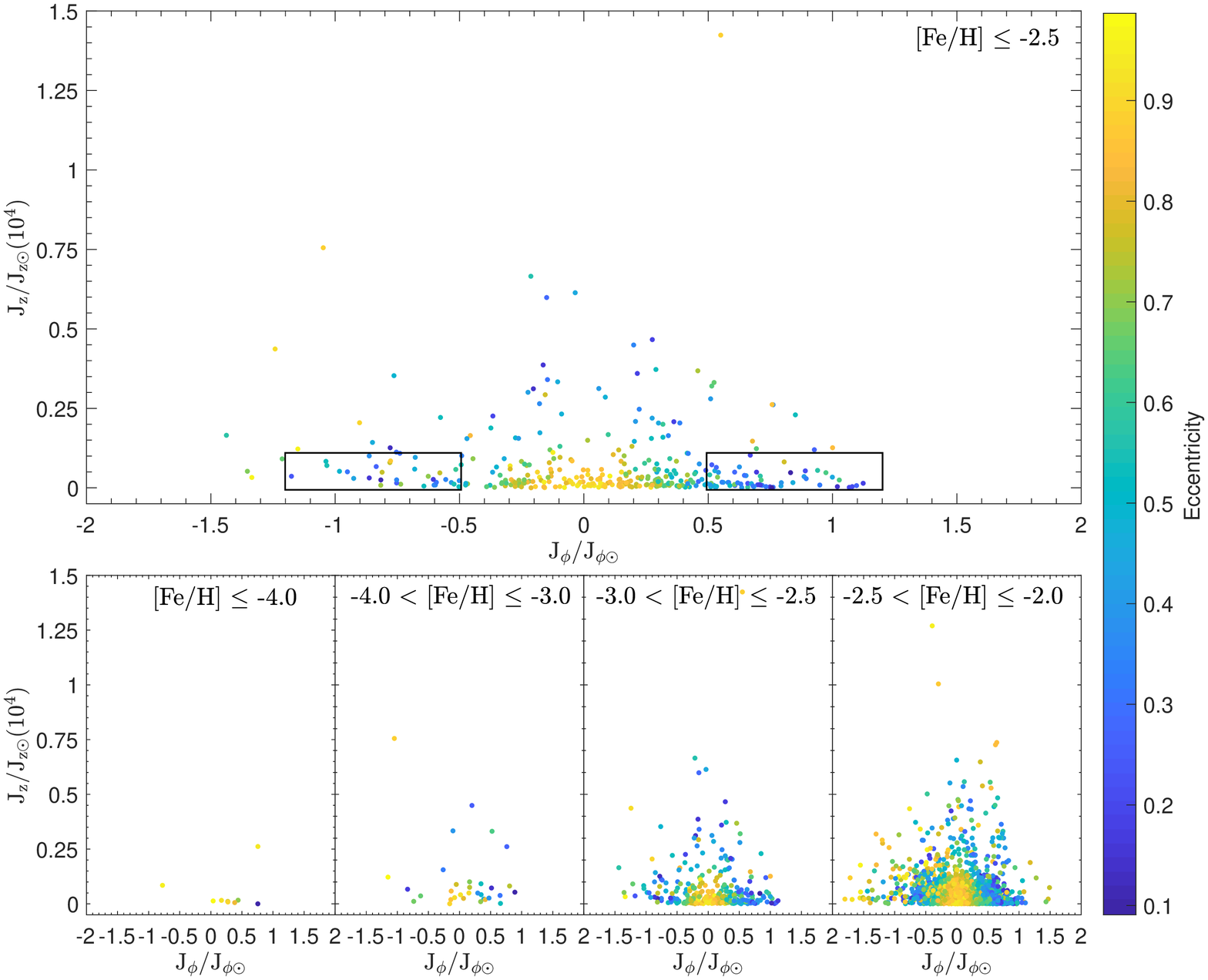}
\caption{Vertical action vs. azimuthal action component colour-coded by eccentricity for the stars with $\delta_{\varpi}/\varpi<0.1$. Top panel: our sample+\citet{Sestito19} stars with $\FeH \leq -2.5$ are shown. Bottom left panel: UMP ($\FeH \leq -4.0$) stars from \citet{Sestito19}. Bottom centre-left panel: stars with $-4.0 < \FeH \leq -3.0$. Bottom centre-right panel:  stars with $-3.0 < \FeH \leq -2.5$. Bottom right panel: stars with $-2.5 < \FeH \leq -2.0$. The sample of stars with $\delta_{\varpi}/\varpi<0.1$ is the $\sim45\%$ of the total.}\label{actionpar}
\end{figure*}

Here in Figure~\ref{actionpar}, we show the same action plot as Figure~\ref{action} but for the stars with $\delta\varpi/\varpi<10\%$ and $\varpi>0 $ mas. It illustrates that our results are similar whether we restrict ourselves only to these stars with the most reliable parallax information.

\bsp	
\label{lastpage}
\end{document}